# MICROALGAE-BASED BIOREMEDIATION OF WATER CONTAMINATED BY PESTICIDES IN PERI-URBAN AGRICULTURAL AREAS

Mª Jesús García-Galán[1*], Luis Simon Monllor-Alcaraz[2], Cristina Postigo[2], Enrica Uggetti[1], Miren López de Alda[2], Joan García[1], Rubén Díez-Montero[1]

[1] GEMMA – Group of Environmental Engineering and Microbiology, Department of Civil and Environmental Engineering, Universitat Politècnica de Catalunya-BarcelonaTech, c/Jordi Girona 1-3, Building D1, E-08034 Barcelona, Spain

[2] Water and Soil Quality Research Group, Department of Environmental Chemistry, Institute of Environmental Assessment and Water Research (IDAEA-CSIC), C/Jordi Girona 18-26, 08034. Barcelona, Spain.

*Corresponding author: María Jesús García Galán

Tel: +34934016204

Email: chus.garcia@upc.edu

## ABSTRACT

The present study evaluated the capacity of a semi-closed, tubular horizontal photobioreactor (PBR) to remove pesticides from agricultural run-off. The study was carried out in July to study its efficiency under the best conditions (highest solar irradiation). A total of 51 pesticides, including 10 transformation products, were selected and investigated based on their consumption rate and environmental relevance. Sixteen of them were detected in the agricultural run-off, and the estimated removal efficiencies ranged from negative values, obtained for 3 compounds, namely terbutryn, diuron, and imidacloprid, to 100%, achieved for 10 compounds. The acidic herbicide MCPA was removed by 88% on average, and the insecticides 2,4-D and diazinon showed variable removals, between 100% and negative values. The environmental risk associated with the compounds still present in the effluent of the PBR was evaluated using hazard quotients (HQs), calculated using the average and highest measured concentrations of the compounds. HQ values > 10 (meaning high risk) were obtained for imidacloprid (21), between 1 and 10 (meaning moderate risk) for 2,4-D (2.8), diazinon (4.6) and terbutryn (1.5), and < 1 (meaning low risk) for the remaining compounds diuron, linuron, and MCPA. The PBR treatment yielded variable removals depending on the compound, similar to conventional wastewater treatment plants. This study provides new data on the capacity of microalgae-based treatment systems to eliminate a wide range of priority pesticides under real/environmental conditions.

## 1. INTRODUCTION

Agricultural and livestock activities are probably the main source of diffuse pollution to both surface and groundwater systems in rural areas (Dolliver and Gupta, 2008; Sabourin et al., 2009; Topp et al., 2008). The increasing population worldwide has led to more intensive animal farming operations and agricultural activities to satisfy food demands, with the consequent increase in the use of veterinary pharmaceuticals in cattle farming activities (especially antibiotics), and inorganic fertilizers and synthetic pesticides in agriculture. Focusing on pesticides, it is estimated that 3 million metric tons of pesticide active ingredients are applied to croplands annually worldwide, and this consumption has increased 15-20 fold in only 30 years (Oerke, 2006; Popp et al., 2013). Their use in agriculture ensures both the quality and the quantity of food production, but also means their entrance in the environment with the subsequent potential risk for soil and water ecosystems, and eventually for human health. Many of these compounds, especially those developed in the last three decades to substitute the toxic and persistent organochlorine pesticides, are highly soluble to reach root systems easily. Therefore, pesticide losses from agricultural fields take place through field run-off and drainage after irrigation or during storm events, reaching surface waters, aquifers, and eventually coastal waters (Ccanccapa et al., 2016; Köck-Schulmeyer et al., 2019; Postigo et al., 2016). These polar pesticides can also be adsorbed or absorbed in different soil layers and sediments (Barbieri et al., 2020; Köck-Schulmeyer et al., 2013a; Palma et al., 2015). Drainage and open irrigation channels can receive a large amount of this agricultural run-off; their diversion into main collectors and towards wastewater treatment plants (WWTPs) require substantial economic investments. Consequently, they usually discharge into rivers, spreading the different pollutants that may indirectly affect a large number of non-target species (Proia et al., 2013; Terzopoulou and Voutsa, 2017). Thus, the equilibrium of river and stream ecosystems can be jeopardized if their capacity to attenuate and neutralize these inputs (dilution, biodegradation, etc.) is overcome. Groundwater systems can also indirectly be affected if polluted rivers feed aquifers by natural infiltration processes (Postigo et al., 2016). Pesticides also reach urban WWTPs due to non-agricultural uses such as grass maintenance (parks, golf courses, etc.), public health protection from plagues (rodents, mosquitos, etc.), urban horticulture and plant nurseries. The removal efficiencies (RE%) obtained during conventional activated sludge (CAS) treatment are usually poor (Köck-Schulmeyer et al., 2013b; Rodriguez-Mozaz et al., 2015). All the research data gathered and risk assessments conducted so far have led to regulate and even ban the use of different pesticides in the EU. In Directive 2013/39/EU, 21 pesticides were considered priority substances and their presence in freshwater ecosystems must be strictly controlled and should not surpass the environmental quality standards (EQS) in surface waters and biota. Several pesticides are also included in the EU Watch List (Commission Implementing decision EC/2018/840), thus being potential candidates to become priority substances in future revisions. Overall, to fulfill those requirements, there is a need to substantially improve the removal capacity of the current wastewater treatment systems.

Presently, great interest is arising in microalgae-based treatment technologies, due to their demonstrated capacity to remove nutrients, organic matter and other pollutants such as heavy metals and also faecal contamination indicators (Abdel-Raouf et al., 2012; García et al., 2006). Microalgae can grow in low quality water such as wastewaters, which still contain high amounts of crucial nutrients for their growth such as $NH_4^+$ or P (from phosphates, $PO_4^{3+}$) and enough inorganic carbon usually as bicarbonate (Markou et al., 2014). They also benefit from the high availability of light in these systems to fixate $CO_2$; and the $O_2$ produced is used by heterotrophic bacterial communities to biodegrade organic matter; likewise, the $CO_2$ from the bacterial respiration can be fixated by microalgae too. Another main advantage of these systems compared to conventional wastewater treatment systems are the low operational and maintenance (O&M) costs. Due to the photosynthetic activity of microalgae, the PBR does not require electromechanical aeration or oxygenation of the mixed liquor (which is the main energetic cost of conventional wastewater treatment) or any chemical input to improve its performance; this advantage is more evident when compared to advanced processes, such as membrane bioreactors or advanced oxidation processes. On the other hand, natural systems require longer retention times and more land surface for their implementation. Overall, in these systems, not only clean water is obtained, but the microalgal biomass produced can be further processed and converted into energy (biogas) (Arashiro et al., 2019; Passos and Ferrer, 2014) and other added-value bioproducts. Indeed, microalgae biomass is being investigated as a potential producer of biofertilizers (Khan et al., 2019), biopolymers (Arias et al., 2020; Rueda et al., 2020), pigments and metabolites (Arashiro et al., 2020; Cuellar-Bermudez et al., 2015), implying a nearly zero-residue process.

Open microalgae systems, usually high rate algal ponds, (HRAPs), have been used for decades, especially in industrial microalgae production due to their lower O&M costs and energy consumption compared to closed systems (Chisti, 2013; Oswald, 1995). However, these systems are exposed to poorer control of environmental parameters (i.e. temperature, salinity, and solar radiation) and they are under a higher risk of contamination (predators, fast-growing heterotrophs). In contrast, closed photobioreactors (PBRs) provide better protection against culture contamination, better control of the operation parameters, less evaporative losses, and higher biomass productions (Chisti, 2007; García-Galán et al., 2018). Yet, closed systems also present some main drawbacks such as higher material and O&M costs compared to open systems, problems with toxic accumulation of dissolved oxygen (DO), and biofouling amongst others. In consequence, the combination of open and closed reactors is probably the most effective configuration for growing algae, overcoming the drawbacks aforementioned to different extents (García-Galán et al., 2018; Uggetti et al., 2018). Overall, microalgae-based wastewater treatment is considered amongst the most environmentally advantageous and least expensive, being ideal for implementation in rural areas. It has been recently demonstrated that microalgae can grow using agricultural run-off as feedstock with good results for both microalgae production and efficient wastewater treatment, removing nutrients and also contaminants of emerging concern (García-Galán et al., 2018).

Within this context, the objective of the present work is to evaluate the capacity of a full-scale semi-closed, tubular horizontal PBR to treat agricultural run-off in the peri-urban agricultural area of Barcelona (Spain), focusing on its capacity to remove a total of 51 medium to highly polar pesticides. The targeted pesticides have been selected based on their inclusion in the EU list of priority substances (Directive 2013/39/EU) and in the EU watch list (Commission Implementing Decision (EU) 2018/840), as well as considering their consumption at Catalonia and European level.

## 2. MATERIALS AND METHODS

### 2.1. Case study

The PBR is located in the experimental campus "Agròpolis" of the Universitat Politècnica de Catalunya-BarcelonaTech (UPC), in Viladecans (Barcelona). Agròpolis is right beside the agricultural area of the Llobregat Delta that belongs to the Baix Llobregat Agrarian Park (Figure 1). This Park was created in 1996 through an EU's LIFE-Environment program, after different municipalities and farmers associations made evident the need to preserve and maintain the management of the surrounding natural and agricultural areas, threatened by the urban and industrial expansion of the city of Barcelona (approximately 10 km away). This park includes the alluvial plains of the Llobregat Delta and the lower valley of the Llobregat River, with approximately 2900 Ha of fruit and vegetable crops (Montasell i Dorda and Callau i Berenguer, 2008).

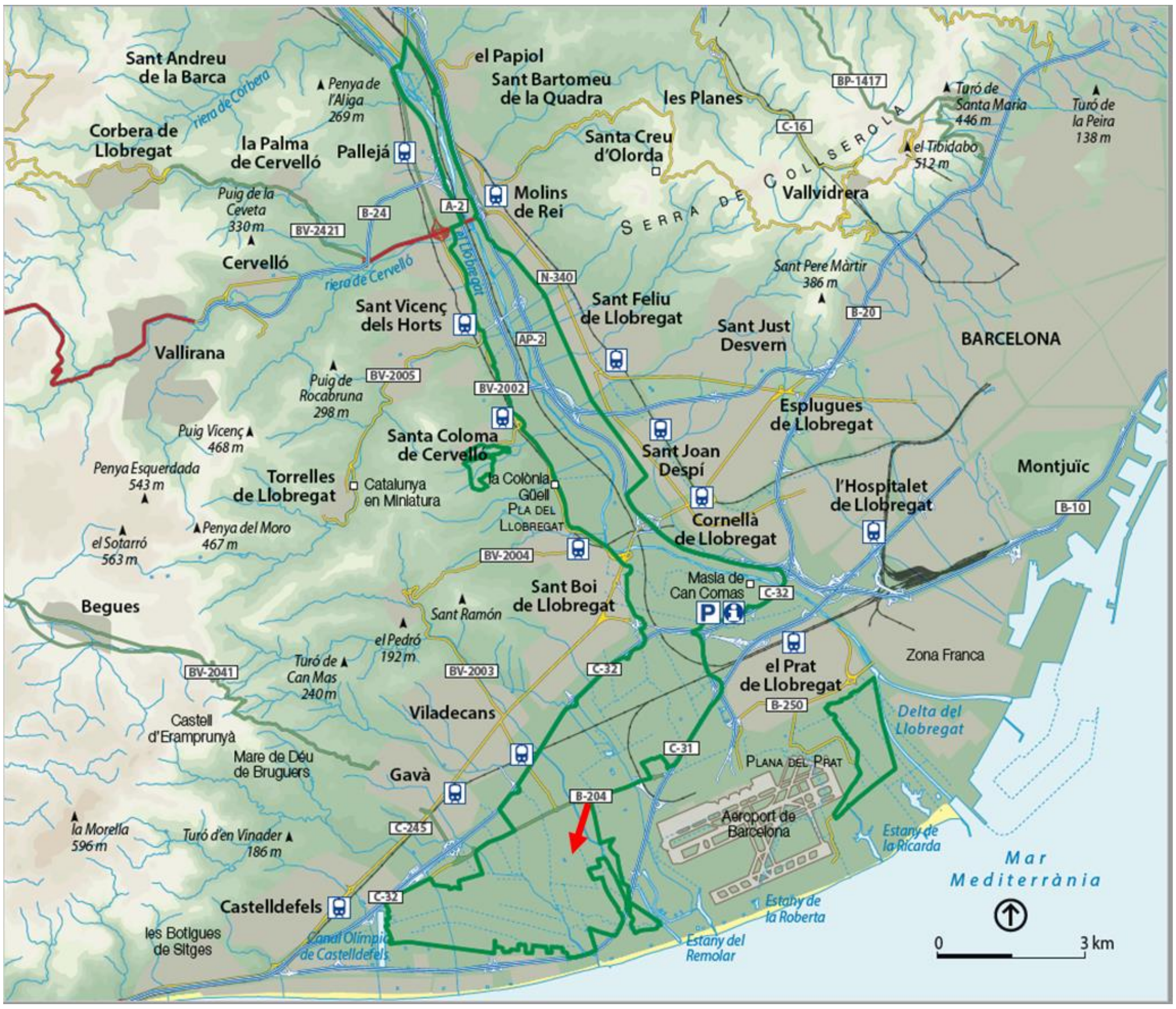

**Figure 1.** Map of the Baix Llobregat Agricultural Park (highlighted in green). Agròpolis (UPC experimental campus) approximated location is pointed by the red arrow.

## 2.2. PBR description and operation

The PBR studied was part of a more complex demonstrative plant implemented within the framework of the European project INCOVER "Innovative Eco-technologies for Resource Recovery from Wastewater" (GA 689242). The main objective of this plant was to generate profitable bioproducts from wastewater and microalgal biomass, within the circular economy and biorefinery paradigms. The configuration of the PBR is depicted in Figure 2.

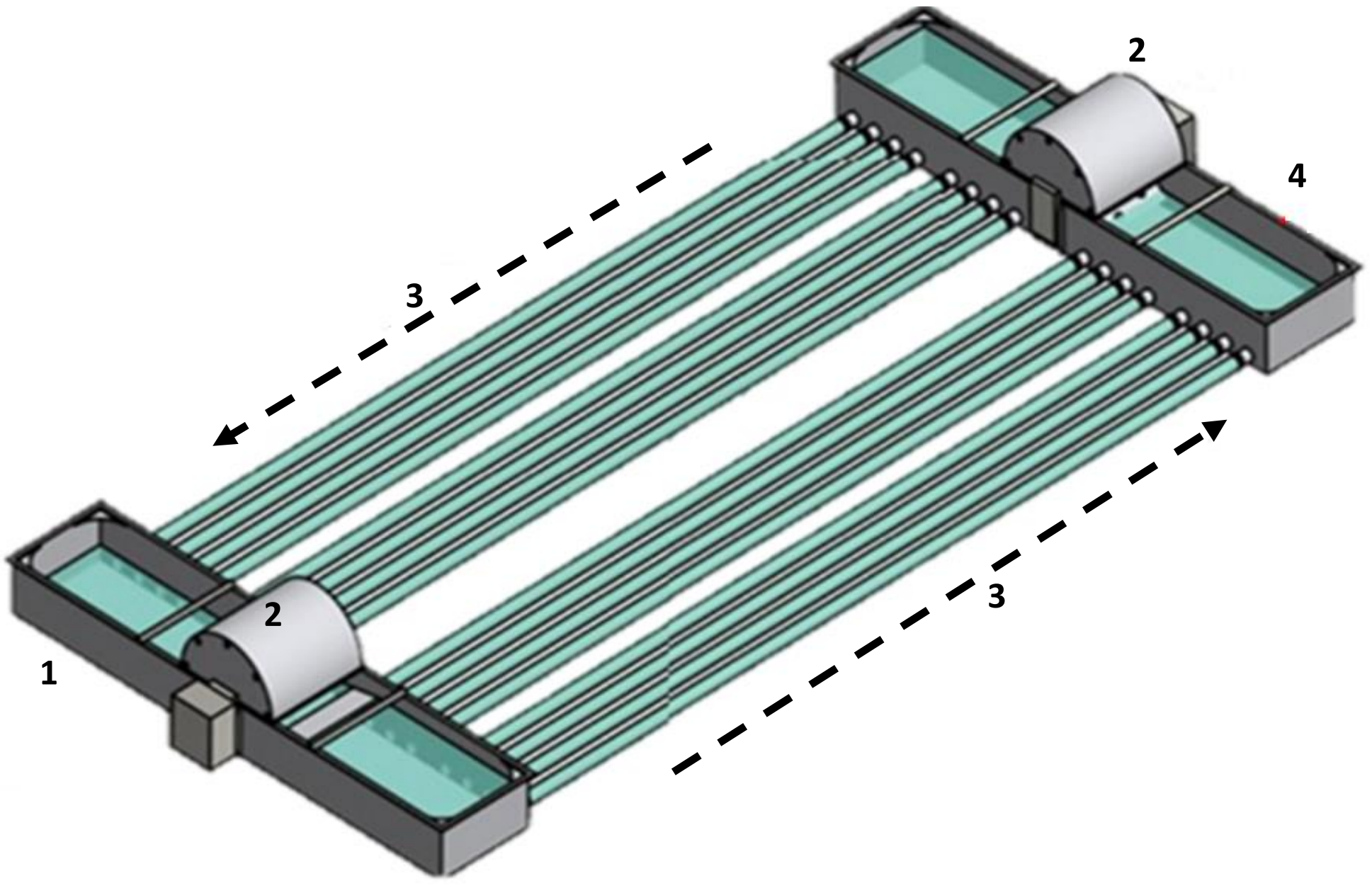

**Figure 2.** Scheme of the hybrid photobioreactor used in this study. 1:inflow from the homogenization tank; 2: paddle wheel; 3: direction of the flow within the tubes; 4: outflow to the storage tanks. Samples were taken in 1 (feeding pipe) and 4 (within the open tank), representing the influent and the effluent of the system, respectively.

Briefly, it consisted of two open tanks connected by 16 horizontal polyethylene tubes (125 mm diameter and 47 m length). The paddlewheels, located in the middle of each open tank, contributed to mix and homogenize the mixed liquor, aiding also to release the excess of dissolved oxygen (DO) that may accumulate within the closed tubes. A dam located after each paddlewheel creates a water level difference that allows the circulation of the mixed liquor from one tank to the other by gravity and through 8 tubes, and back through the other 8 tubes. The PBR had a useful volume of 11.7 $m^3$. It was inoculated with 10 L of a mixed culture of microalgae from a pilot HRAP treating urban wastewater. The agricultural run-off was collected from an irrigation channel and pumped to the pilot plant, where it was mixed with partially treated domestic wastewater from a septic tank (mixing ratio 7:1, v:v (run-off: domestic wastewater)). The mixture was homogenized in a 10 $m^3$ tank with constant stirring to avoid suspended solids sedimentation and was fed to the PBR daily, under a hydraulic retention time (HRT) regime of 5 d (2.3 $m^3$ of daily feedstock). Technical and operational parameters are summarized in Table 1. Further information on the design and operation of the PBR is given elsewhere (García et al., 2018; Uggetti et al., 2018).

**Table 1**. Technical characteristics and operational parameters of the studied PBR

| Parameter | Value |
|---|---|
| **Total volume ($m^3$)** | 11.7 |
| **Tank volume ($m^3$)** | 1.25 |
| **Tube lenght and diameter (m)** | 47/0.125 |
| **Volume in the tubes ($m^3$)** | 9.2 |
| **Flow rate ($m^3$/d)** | 2.3 |
| **HRT (d)** | 5 |

## 2.3. Chemicals and reagents

Standards of the 51 target pesticides and 45 isotopically-labeled pesticides used as surrogate standards were purchased from Fluka (Sigma-Aldrich, Steinheim, Germany) or Dr. Ehrenstorfer (LGC Standards, Teddington, UK). The list of target analytes includes eight triazines (atrazine, simazine, cyanazine, cybutryn, terbuthylazine, terbutryn, and the atrazine transformation products (TPs) desethylatrazine and deisopropylatrazine); four phenylureas (diuron, isoproturon, linuron, and chlortoluron); twelve organophosphates (azinphos ethyl, azinphos-methyl and its TP azinphos-methyl oxon, chlorfenvinphos, dichlorvos, diazinon, dimethoate, chlorpyrifos, malathion, malaoxon, fenitrothion and its TP fenitrothion oxon); six organothiophosphates (fenthion and six of its TPs, namely fenthion oxon, fenthion sulfone, fenthion sulfoxide, fenthion oxon sulfoxide, fenthion oxon sulfone, and fenthion oxon sulfoxide); five neonicotinoids (acetamiprid, clothianidin, imidacloprid, thiacloprid, and thiamethoxam); two chloroacetanilides (alachlor and metolachlor); four acidic herbicides (mecoprop, 2,4 D, bentazone, and MCPA). Other individual analytes were molinate, bromoxynil, diflufenican, methiocarb, oxyfluorfen, thifensulfuron methyl, pendimethalin, quinoxyfen, propanil, fluroxypir, and oxadiazon. More info on the target analytes is given in Table 2. The use of some of the target pesticides, i.e., the triazines atrazine, simazine, cyanazine, and terbutryn, the organophosphates chlorfenvinphos, chlorpyrifos, diazinon and fenitrothion, the anilide propanil, and the chloroacetanilides alachlor and metolachlor is banned in the EU. Eight of the target pesticides, namely atrazine, simazine, alachlor, chlorfenvinphos, terbutryn, isoproturon, and irgarol, are considered priority substances in surface waters following Directive 2013/39/EU, and six are in the EU Watch List (methiocarb, and the neonicotinoids acetamiprid, clothianidin, imidacloprid, thiacloprid, and thiamethoxam).

Mix standard solutions were prepared and used as spiking solutions to prepare eleven aqueous calibration solutions in the range 0.1 – 1000 ng $L^{-1}$ that contained a fixed amount of the surrogate standard mixture (200 ng $L^{-1}$). Pesticides-grade solvents MeOH, acetonitrile (ACN), and LC-grade water used for pesticide analysis were supplied by Merck (Darmstadt, Germany).

**Table 2.** List of the physical and chemical properties of the targeted pesticides. Compounds in italics are metabolites or TPs. MW: molecular weight; S: solubility.

| | | **CAS NUMBER** | **FORMULA** | **MW** | **S (mg $L^{-1}$)** | **$K_{oc}$ (mg $g^{-1}$)** | **$K_{ow}$ (logP)** | **pKa** |
|---|---|---|---|---|---|---|---|---|
| **Acidic** | **2,4-D** | 94-75-7 | $C_8H_6Cl_2O_3$ | 221.04 | 24300 | 39 | -0.82 | 3.4 |
| | ***MCPA*** | 94-74-6 | $C_9H_9ClO_3$ | 200.62 | 29390 | 29 | -0.81 | 3.73 |
| **Organo-phosphates** | **Diazinon** | 333-41-5 | $C_{12}H_{21}N_2O_3PS$ | 304.35 | 60 | 609 | 3.69 | 2.6 |
| | **Dimethoate** | 60-51-5 | $C_5H_{12}NO_3PS_2$ | 229.26 | 25900 | 25 | 0.75 | - |
| | **Fenitrothion** | 122-14-5 | $C_9H_{12}NO_5PS$ | 277.23 | 19 | 2000 | 3.32 | - |
| | **Fenitrothion oxon** | 2255-17-6 | $C_9H_{12}NO_6P$ | 261.17 | 301 | 21 | 1.69 | - |
| | **Chlorfenvinphos** | 470-90-6 | $C_{12}H_{14}Cl_3O_4P$ | 359.6 | 145 | 680 | 3.8 | - |
| | **Chlorpyrifos** | 2921-88-2 | $C_9H_{11}Cl_3NO_3PS$ | 350.58 | 1.05 | 5509 | 4.7 | - |
| | **Malathion** | 121-75-5 | $C_{10}H_{19}O_6PS_2$ | 330.36 | 148 | 1800 | 2.75 | - |
| | **Malaoxon** | 1634-78-2 | $C_{10}H_{19}O_7PS$ | 314.29 | 7500 | 4650 | 0.52 | - |
| | **Azinphos-Methyl** | 86-50-0 | $C_{10}H_{12}N_3O_3PS_2$ | 317.32 | 28 | 1112 | 2.96 | 5 |
| | **Azinphos-Methyl-Oxon** | 961-22-8 | $C_{10}H_{12}N_3O_4PS$ | 301.26 | 2604* | 10* | 0.77 | - |
| | **Azinphos-Ethyl** | 2642-71-9 | $C_{12}H_{16}N_3O_3PS_2$ | 345.38 | 4.5 | 1500 | 3.18 | 1.4 |
| | **Dichlorvos** | 62-73-7 | $C_4H_7C_{12}O_4P$ | 220.98 | 18000 | 50 | 1.9 | - |
| **Organ thiophosphates** | **Fenthion** | 55-38-9 | $C_{10}H_{15}O_3PS_2$ | 278.33 | 4.2 | 1500 | 4.84 | - |
| | ***Fenthion oxon*** | 6552-12-1 | $C_{10}H_{15}O_4PS$ | 262.26 | 213.5 | 57 | 2.31 | - |
| | ***Fenthion sulfone*** | 3761-42-0 | $C_{10}H_{15}O_5PS_2$ | 310.33 | 190.4 | 542 | 2.05 | - |
| | ***Fenthion sulfoxide*** | 3761-41-9 | $C_{10}H_{15}O_4PS_2$ | 294.33 | 3.72 | 466 | 1.92 | - |
| | ***Fenthion oxon sulfone*** | 14086-35-2 | $C_{10}H_{15}O_6PS$ | 294.03 | 7602 | 13 | 0.28 | - |
| | ***Fenthion oxon sulfoxide*** | 6552-13-2 | $C_{10}H_{15}O_5PS$ | 278.26 | 1222 | 11 | 0.15 | - |
| Triazines | **Cyanazine** | 21725-46-2 | $C_9H_{13}ClN_6$ | 240.69 | 171 | 190 | 2.1 | 12.9 |
| | ***Desethylatrazine*** | 6190-65-4 | $C_6H_{10}ClN_5$ | 187.63 | 2700 | 110 | 1.51 | - |
| | ***Deisopropylatrazine*** | 1007-28-9 | $C_5H_8ClN_5$ | 173.6 | 980 | 130 | 1.15 | - |
| | **Terbuthylazine** | 5915-41-3 | $C_9H_{16}ClN_5$ | 229.71 | 6.6 | 329 | 3.4 | 1.9 |
| | **Terbutryn** | 886-50-0 | $C_{10}H_{19}N_5S$ | 241.36 | 25 | 2432 | 3.66 | 4.3 |
| | **Cybutryn** | 28159-98-0 | $C_{11}H_{19}N_5S$ | 253.37 | 7 | 1569 | | |
| | **Simazine*** | 122-34-9 | $C_7H_{12}ClN_5$ | 201.66 | 5 | 130 | 2.3 | 1.62 |
| | **Atrazine*** | 1912-24-9 | $C_8H_{14}ClN_5$ | 215.68 | 35 | 100 | 2.7 | 1.7 |
| | **Irgarol** | 28159-98-0 | $C_{11}H_{19}N_5S$ | 253.37 | 7 | 1569 | 3.95 | - |
| Phenyl ureas | **Linuron** | 330-55-2 | $C_9H_{10}Cl_2N_2O_2$ | 249.09 | 63.8 | 843 | 3 | - |
| | **Chlortoluron** | 15545-48-9 | $C_{10}H_{13}ClN_2O$ | 212.68 | 74 | 196 | 2.5 | - |
| | **Diuron*** | 330-54-1 | $C_9H_{10}Cl_2N_2O$ | 233.09 | 35.6 | 813 | 2.87 | - |
| | **Isoproturon*** | 34123-59-6 | $C_{12}H_{18}N_2O$ | 206.28 | 70.2 | 251 | 2.5 | - |
| Neonicoti-noids | **Acetamiprid ** ** | 135410-20-7 | $C_{10}H_{11}ClN_4$ | 222.67 | 2950 | 200 | 0.8 | 0.7 |
| | **Clothianidin ** ** | 210880-92-5 | $C_6H_8ClN_5O_2S$ | 249.68 | 340 | 123 | 0.9 | 11.1 |
| | **Imidacloprid ** ** | 138261-41-3 | $C_9H_{10}ClN_5O_2$ | 255.66 | 610 | 6719 | 0.57 | - |
| | **Thiacloprid ** ** | 111988-49-9 | $C_{10}H_9ClN_4S$ | 252.72 | 184 | 615 | 1.26 | - |
| | **Thiamethoxam *** | 153719-23-4 | $C_8H_{10}ClN_5O_3S$ | 291.71 | 4100 | 56.2 | <3 | - |
| Chloro-acetanilides | **Alachlor*** | 15972-60-8 | $C_{14}H_{20}ClNO_2$ | 269.77 | 240 | 335 | 3.09 | 0.62 |
| | **Metolachlor** | 51218-45-2 | $C_{15}H_{22}ClNO_2$ | 283.8 | 530 | 120 | 3.4 | - |
| Thiocarbamates | **Molinate** | 2212-67-1 | $C_9H_{17}NOS$ | 187.3 | 1100 | 190 | 2.86 | |
| Benzothiadiazines | **Bentazone** | 25057-89-0 | $C_{10}H_{12}N_2O_3S$ | 240.3 | 7112 | 55 | -0.42 | 3.51 |
| Other | **Bromoxinil** | 1689-84-5 | $C_7H_3Br_2NO$ | 276.9 | 38000 | 302 | 0.27 | 3.86 |
| | **Diflufenican** | 83164-33-4 | $C_{19}H_{11}F_5N_2O_2$ | 394.29 | 0.05 | 3,19* | 4.2 | - |
| | **Methiocarb ** ** | 2032-65-7 | $C_{11}H_{15}NO_2S$ | 225.31 | 27 | 182 | 3.18 | - |
| | ***Thifensulfuron Methyl*** | 79277-27-3 | $C_{12}H_{13}N_5O_6S_2$ | 387.39 | 54.1 | 28 | -1.65 | 4 |
| | **Pendimethalin** | 40487-42-1 | $C_{13}H_{19}N_3O_4$ | 281.31 | 0.33 | 17491 | 5.4 | 2.8 |
| | **Quinoxyfen** | 124495-18-7 | $C_{15}H_8Cl_2FNO$ | 308.13 | 0.05 | 23 | 4.66 | - |
| | **Propanil** | 709-98-8 | $C_9H_9Cl_2NO$ | 218.08 | 95 | 149 | 2.29 | 19.1 |
| | **Fluroxypir** | 69377-81-7 | $C_7H_5Cl_2FN_2O_3$ | 255.03 | 6500 | - | 0.04 | 2.94 |
| | **Mecoprop** | 7085-19-0 | $C_{10}H_{11}ClO_3$ | 214.65 | 250000 | 47 | -0.19 | 3.11 |
| | **Oxadiazon** | 19666-30-9 | $C_{15}H_{18}Cl_2N_2O_3$ | 345.2 | 0.57 | 3200 | 5.33 | - |

Information obtained from the pesticides properties database (https://sitem.herts.ac.uk/aeru/footprint/es/).

*: Included in the Directive 2013/39/EC as *priority substances;* **: Included in the *watch list* of the Commission implementing decision (EU) 2015/495;

## 2.4. Sample collection

Sampling was performed during two consecutive weeks in July. Influent water (agricultural run-off) and mixed liquor (PBR effluent) from one of the open tanks of the PBR were taken three days each week (n=12), always at 10.00 am. The samples for pesticide analysis were collected in amber polyethylene terephthalate (PET) bottles and transported under cool conditions to the laboratory. They were immediately fortified with the mixture of deuterated surrogate standards at 200 ng $L^{-1}$, centrifuged to remove suspended particles (3500 rpm for 10 min at room temperature) (5810R centrifuge, Eppendorf Ibérica, Spain), and stored at -20 °C in the dark until analysis. Samples for water chemical characterization were taken in PVC bottles and analyzed in the laboratory upon arrival.

## 2.5. Analytical methodologies

### 2.5.1. Water samples characterization

On-site measurements of water temperature, dissolved oxygen (DO), and pH were performed using online sensors submerged in one of the open tanks and connected to a Multimeter 44 (Hatch Lange SL., Spain). Data were registered every 5 s and recorded and stored each 60 s by a datalogger (Campbell Scientific Inc., USA). Influent samples were analyzed for these parameters using portables devices for DO and temperature (EcoScan DO 6, ThermoFisher Scientific, USA) and pH (Crison 506, Spain). Both influent and mixed liquor samples were also analyzed on turbidity (Hanna HI 93703, USA), total suspended solids (TSS), volatile suspended solids (VSS), alkalinity and total and soluble chemical oxygen demand (COD and CODs) following Standard Methods (APHA-AWWA-WEF, 2012); $NH_4^+$-N was analyzed according to Solórzano method (Solórzano, 1969). The ions $NO_2^-$-N, $NO_3^-$-N and $PO_4^{3-}$-P were measured by ion chromatography (ICS-1000, Dionex Corporation, USA). Total carbon (TC), total phosphorus (TP) and total nitrogen (TN) were measured by a multi N/C 2100S (Analytik Jena, Germany). All the analyses were performed in triplicate. Mixed liquor samples were regularly examined under an optic microscope (Motic, China) for qualitative evaluation of microalgae populations employing taxonomic books and databases for their identification (Bourrelly, 1990; Palmer, 1962).

Average biomass productivities (gVSS $m^{-2}\cdot d^{-1}$) in the PBR were calculated based on the VSS concentration in the mixed liquor of both systems, using equation [1]:

$$Biomass\ productivity = \frac{VSS\ (Q - Q_E + Q_P)}{A} \qquad [1]$$

where $VSS$ is the volatile suspended solids concentration of the PBR mixed liquor (g VSS $L^{-1}$); $Q$ is the wastewater flow rate (L $d^{-1}$); $Q_E$ is the evaporation rate (L $d^{-1}$); $Q_P$ is the precipitation rate (L $d^{-1}$); and $A$ is the surface area of the system (for the PBR, it was calculated including both tanks surfaces and half of the surface of the 16 tubes). The evaporation rate was calculated using [2]:

$$Q_E = E_p\, A \qquad [2]$$

where $E_p$ is the potential evaporation (mm $d^{-1}$), calculated using equation [3] (Fisher and Pringle III, 2013).

$$E_p = a\ \frac{T_a}{(T_a + 15)}(R + 50) \qquad [3]$$

where $a$ is a dimensionless coefficient which varies depending on the sampling frequency (0.0133 for daily samples); $R$ is the average solar radiation in a day (MJ $m^{-2}$), and $T_a$ is the average air temperature (°C). Meteorological data were obtained from the weather stations network located in Barcelona and the metropolitan area (*www.meteo.cat*).

### 2.5.2. Analysis of pesticides

Determination of pesticide concentrations in the water samples was achieved following a fully automated methodology consisting of on-line solid-phase extraction and liquid chromatography-tandem mass spectrometry analysis (on-line SPE-LC-MS/MS) (Palma et al., 2014). Briefly, 5 mL of the samples (as well as calibration solutions and blanks) were extracted onto a previously conditioned (1 mL of ACN and 1 mL of LC-grade water) CHROspe cartridges (Axel Semrau GmbH & Co. KG) using an automated SPE sample processor Prospekt-2 (Spark Holland, Emmen, The Netherlands). After sample loading (1 mL $min^{-1}$), the SPE cartridges were washed with 1 mL of LC-grade water (5 mL $min^{-1}$) and eluted directly onto the LC column. Analyte detection was carried out with a TQD triple-quadrupole mass spectrometer, equipped with an electrospray (ESI) interface (Waters, Milford, MA, USA). MS/MS detection was performed in the selected reaction monitoring (SRM) mode, recording two SRM transitions per target analyte. The ESI interface was operated in both positive and negative ionization modes, allowing for the simultaneous analysis of all target pesticides in one single analytical run (43 compounds were analyzed in ESI+ and 8 in ESI-). Detailed information on the SRM transitions and ionization conditions for each selected analyte is given elsewhere (Barbieri et al., 2019). Regarding the method sensitivity, limits of detection (LOD) and quantification (LOQ) were calculated experimentally as the minimum detectable amount of analyte with a signal-to-noise ratio of 3 and 10, respectively.

### 2.6. Environmental risk assessment

The environmental risk evaluation of the pesticides present in the PBR effluent samples was based on the calculation of hazard quotients (HQs), which estimate the potential adverse effects of the pesticide concentrations detected in the environmental matrix studied on non-target organisms. This quotient is calculated as the ratio between the measured environmental concentration (MEC) and the predicted no-effect concentration (PNEC), the concentration at which no toxic effects are expected in aquatic organisms following the equation [4]:

$$HQ = \frac{MEC}{PNEC}$$

The PNEC values were obtained from the NORMAN Ecotoxicology Database (https://bit.ly/2Cm4zOE). When HQ>1, a potential environmental impact could be expected and should be investigated in further detail. Eventually and to evaluate the overall ecotoxicity risk of the PBR effluent samples posed by the various pesticides present in them, toxic units (TU) were generated for each sample as the sum of all HQs calculated for each pesticide detected, on each day of sampling.

## 3. RESULTS AND DISCUSSION

### 3.1. Overall performance of the PBR

The operational performance of the PBR in terms of removal of conventional physical-chemical parameters is summarized in Table 3. The VSS/TSS ratio in the PBR was 74%, which agrees with values typically observed in microalgae-based systems, which are generally > 70% (Arashiro et al., 2019; Santiago et al., 2017). The pH values registered were higher than 8, which is considered the optimum value to promote microalgae growth (Arias et al., 2017). This increase is due to the high photosynthetic activity during the summer season, which implies the assimilation of protons from the media during $CO_2$ fixation (Markou et al., 2014). Average soluble COD values increased by 14% after PBR treatment, which is probably due to the partial release of carbon fixated during photosynthesis in the form of dissolved organic carbon. This pattern has been previously observed in different closed systems (García-Galán et al., 2018; García et al., 2006) and, in consequence, the total carbon (TC) concentration is also higher in the PBR effluent. Arbib et al., (2013) also obtained these higher COD concentrations in the effluent of their closed PBR than in the influent and indicated that this carbon release could represent a 5-30% of the fixed carbon. On the other hand, the low biodegradability potential of the agricultural run-off (low organic matter concentration) and the limited carbon availability can affect the algal growth and the whole efficiency of the system. Indeed, COD removal rates reported in open systems (60%-70%) are usually higher than those in closed systems (García-Galán et al., 2020; Arashiro et al., 2019; Young et al., 2017). Regarding $N\text{-}NH_4^+$, a removal of 83% was obtained, with an initial concentration in the influent of 4.5 mg [$N\text{-}NH_4^+$] $L^{-1}$. In microalgae-based systems, $N\text{-}NH_4^+$ is the preferred form of nitrogen by microalgae, followed by $N\text{-}NO_3^-$ (Maestrini, 1982; Oliver and Ganf, 2002; Ruiz-Marin et al., 2010). Furthermore, the high photosynthetic activity during summer increases the pH and favors $N\text{-}NH_4^+$ volatilization as $N\text{-}NH_3$, as well as its partial nitrification to $N\text{-}NO_3^-$ (Van Den Hende et al., 2016). Thus, the removal of $N\text{-}NO_3^-$ was moderate (54%).

**Table 3.** Physical-chemical characterization of the agricultural run-off (influent wastewater) and PBR effluent.

| Parameter | Agricultural run-off | $PBR_{eff}$ | Removal efficiency |
|---|---|---|---|
| **pH** | 8.3 ± 0.3 | 9.1 ± 1.0 | NA |
| **DO (mg $L^{-1}$)** | - | 8.97 ± 0.86 | NA |
| **Temperature (°C)** | 24.18 ± 2.1 | 24.87 ± 1.6 | NA |
| **TSS (mg $L^{-1}$)** | 73.70 ± 58.81 | 291.18 ± 200.91 | - |
| **VSS (mg $L^{-1}$)** | 20.43 ± 13.85 | 215.35 ± 124.95 | - |
| **CODs (mg$O_2$ $L^{-1}$)** | 92.50± 50.06 | 107.64 ± 81.06 | - |
| **TC (mg $L^{-1}$)** | 162.0 ± 19.9 | 246.3 ± 34.4 | - |
| **$N\text{-}NH_4^+$ (mg $L^{-1}$)** | 4.4 ± 1.5 | 0.3 ± 0.5 | 93.2 |
| **$N\text{-}NO_2^-$ (mg $L^{-1}$)** | 1.1 ± 1.0 | 2.2 ± 3.4 | - |
| **$N\text{-}NO_3^-$ (mg $L^{-1}$)** | 9.3 ± 1.8 | 4.3 ± 5.3 | 53.8 |
| **$P\text{-}PO_4^{3-}$ (mg $L^{-1}$)** | 1.6 ± 1.0 | 0.0 ± 0.0 | 100 |

NA: not applicable

The biomass productivity obtained in the PBR was 6.9 g VSS $m^{-2}$ $d^{-1}$ on average (Figure 3). Arbib et al. (2013) obtained productivities between 4.4 g SS $m^{-2}$ $d^{-1}$ and 8.26 g SS $m^{-2}$ $d^{-1}$ in a small scale PBR (380 L). However, values between 20-40 g VSS $m^{-2}$ $d^{-1}$ are considered typical in closed systems (García-Galán et al., 2018). As discussed above, the carbon limitation of the agricultural run-off probably affected the growth rate of the biomass, as well as the low nutrients concentration observed ($N\text{-}NH_4^+$, Total Inorganic Nitrogen (TIN) and $P\text{-}PO_4^{3-}$). It should be mentioned that biofilm was developed in the inner wall of the tubes. Due to the shearing stress produced by the flow inside the tubes, biofilm detachment also took place. This detachment, combined with regular maintenance, made it possible to keep a balance between biofilm formation and detachment. Anyway, the biofilm could have partially hindered the penetration of sunlight into the mixed liquor, affecting the growth of microalgae and the overall system efficiency.

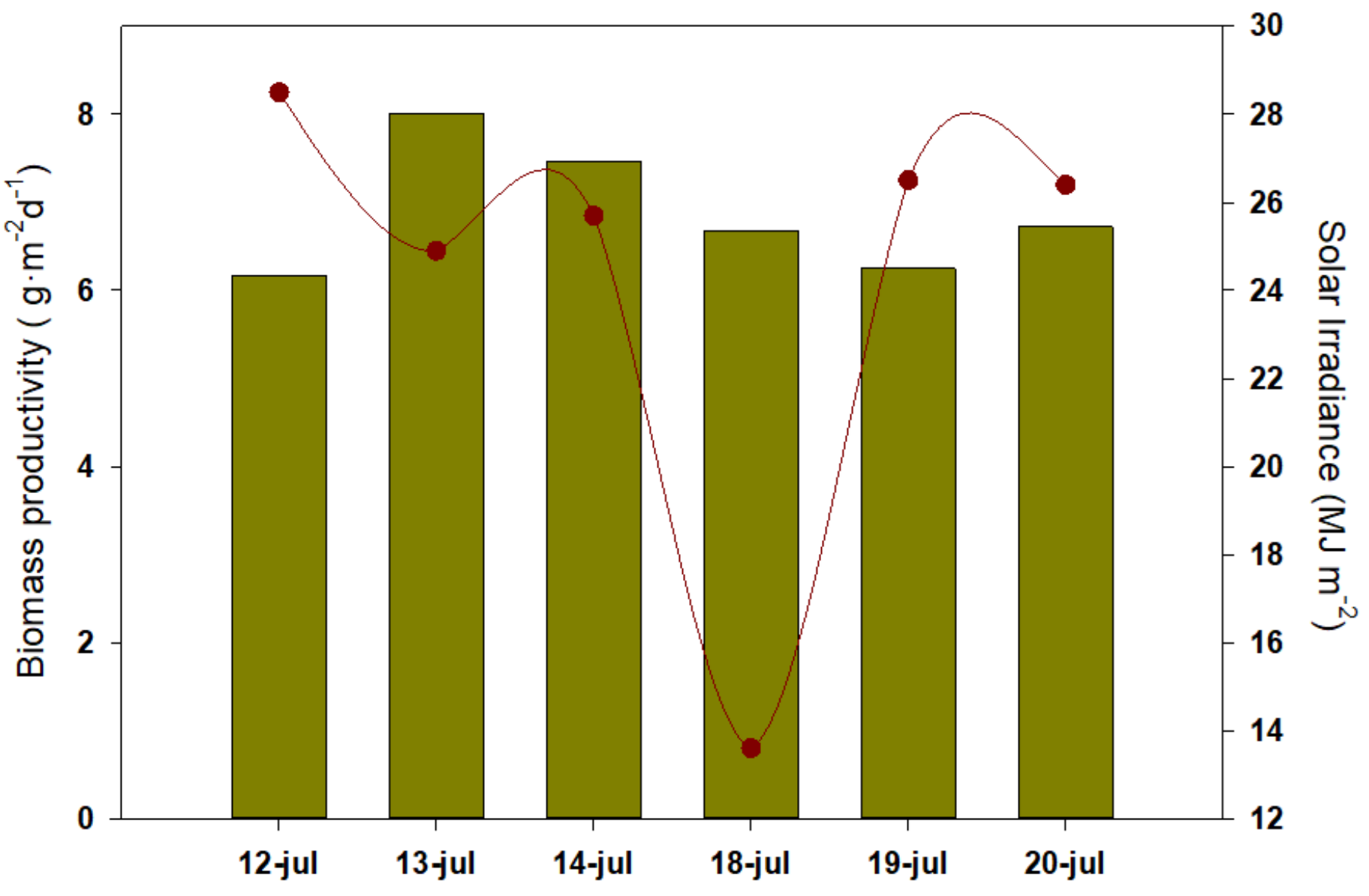

**Figure 3.** Biomass productivity (bars) and average global solar irradiance (red dots) on the PBR studied

The cyanobacteria *Synechocystis* sp was the most abundant species detected in the PBR (Figure 4). Its predominance could be attributed to the abundance of the nutrients in the mixed liquor. In a previous study, Arias et al., (2018) indicated that a concentration of TN < 11.72 mg N L $d^{-1}$, (similar to those found in the PBR of the present work) would favor the prevalence of cyanobacteria.

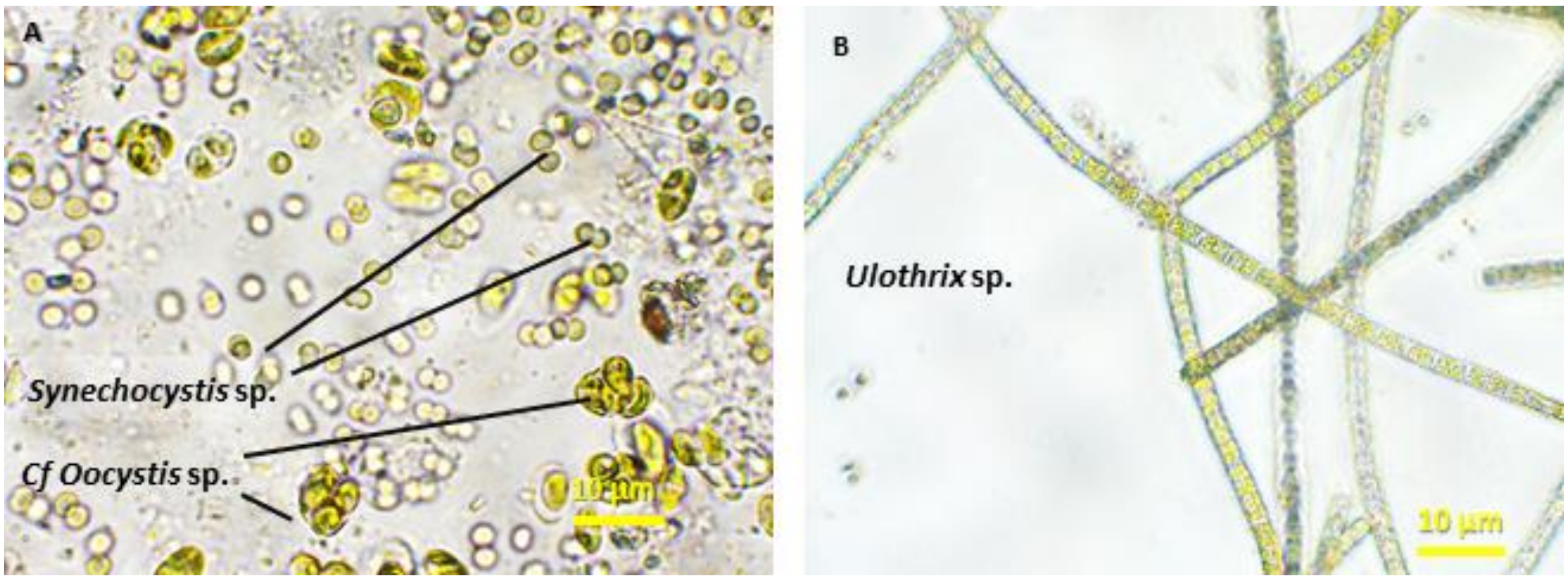

**Figure 4.** Microscope images of mixed liquor of the hybrid PBR (A-B), observed in bright light microscopy (x1000).

### 3.2. Occurrence of pesticides in agricultural run-off

Of the 51 targeted pesticides, 16 of them were detected in the agricultural run-off samples entering the PBR (Table 4). None of the individual concentrations of the pesticides detected were higher than the threshold of concern for microalgae registered in the Pesticide Properties Database (PPDB) (developed by the Agriculture & Environment Research Unit (AERU), University of Hertfordshire). The well-functioning of the PBR, in terms of microalgae mortality, was therefore ensured. The cumulative concentrations of pesticides for each day of sampling are shown in Table 4. Method LODs and LOQs are given for the detected compounds in Table 5 of the SI. LODs ranged from 0.05 to 19 ng $L^{-1}$ and LOQs from 0.15 to 63 ng $L^{-1}$, both for diazinon and desisopropyl atrazine, respectively. The acidic herbicide MCPA was detected at the highest concentration in 5 out of the 6 days of sampling, with levels ranging from 45 ng $L^{-1}$ to 392 ng $L^{-1}$. MPCA and other carboxylic acid herbicides (such as 2,4-D) are commonly used to control the spread of annual and perennial broadleaved weeds on cereal fields, lawns, and vineyards (Kersten et al., 2014). Both are highly soluble and their persistence in soils and sediments is low. MCPA was also detected in all the surface water samples taken in a survey carried out in the Fraser Valley (Canada), highly impacted by agriculture activities (Woudneh et al., 2007). Locally, it was also present in different drainage channels in the Ebro River delta (NE Spain), with maximum and average concentrations of 4197 ng $L^{-1}$ and 356 ng $L^{-1}$, respectively (Köck et al., 2010). It is also frequently detected in urban sites, but at lower concentrations (Köck-Schulmeyer et al., 2013b; Rodriguez-Mozaz et al., 2015). The second acidic pesticide detected, 2,4-D, was present only two days of sampling at 9.7 and 23 ng $L^{-1}$. 2,4-D is one of the most widely used herbicides in the world, and it has been detected in different surface waters all over Europe (Loos et al., 2010, 2009), with maximum levels detected up to 50 ng $L^{-1}$ near Budapest and 20-10 ng $L^{-1}$ in the Austrian part of the Danube River basin (it was detected in 94% of the samples). It has also been detected in urban wastewaters at concentrations in the range of 62–207 ng $L^{-1}$ in Galicia (NW Spain) (Rodil et al., 2012), and average and maximum levels of 86 and 442 ng $L^{-1}$, respectively, in Catalonia (Spain) (Köck-Schulmeyer et al., 2013b).

**Table 4.** Frequency of detection (FD%), maximum and average concentration of individual pesticides detected in the agricultural run-off.

| | | Pesticide | FD (%) | *Max. (ng $L^{-1}$) | Average (ng $L^{-1}$) | Previously reported concentrations (ng $L^{-1}$): Surface waters | Previously reported concentrations (ng $L^{-1}$): Urban waters |
|---|---|---|---|---|---|---|---|
| **HERBICIDES** | **Acidic** | **2,4- D** | 33.3 | 23.2 | 16.4 | 13-14[a], 22[b], 0.5-166.2[c], 1.8-30.9[d], 0-52[e], 5-31[f] | 62-207[e] 47 (inf)[g], 88.5 (inf)[h], 42.9 (eff)[h] |
| | | **MCPA** | 83.3 | 392.3 | 134 ± 30 | 13.1-64.9[c], 0.4-317[d], 76-679[f] | 18-40 (inf)[g] 7.6 (inf)[h], 12.2 (eff)[h], |
| | **Chloro-acetanilides** | **Alachlor** | 16.7 | - | 6.16 | 4.3-11.1[i], 2.2-17[j], 0.5-34.4[c], 0.7-0.9[r] | 2.59 (inf)[h] |
| | **Phenyl Ureas** | **Diuron** | 100 | 60.6 | 41.8 ± 11 | 2-3[a], 41[b], 0.5-92.9[c], 2.4-818[i], 0.4-99.7[j], 64-239[k], 9.5[l], 2.6-150.9[m], 10.5-36[n], 3.5-159.5[o], 0.7-2.3[q], 2.6-24.7[r] | 75-101 (eff)[g] 93 (inf)[h] 127 (eff)[h] |
| | | **Linuron** | 16.7 | - | 13.2 | 1.11-2.5[c], 0.9-1.2[r] | 36 (inf)[g] |
| | **Triazines** | **Cybutryne (irgarol)** | 16.7 | - | 14.2 | | |
| | | **Desisopropyl atrazine (DIA)** | 16.7 | - | 64.2 | 1.9-12.6[c], 0.1-14.4[j], 1.2[l], 25-62[o], 3.9-8.9[r] | 13.7 (inf)[h] 38.8 (eff)[h] |
| | | **Terbuthylazine** | 16.7 | - | 11.3 | 11-14[a], 9[b], 15.9-254.4[c], 4.2-81.5[i], 0.1-21.9[j], 12[l], 185.6-44.6[n], 0.11-10.1[m], 0.3-1[q], 30.9-34.8[r] | 20.6 (infl)[h] 20 (eff)[h] |
| | | **Terbutryn** | 100 | 70 | 33.7 ± 19 | 0.9-30.5[m], 1.04-23.4[o] | |
| **PLAGUICIDES** | **Organo-phosphates** | **Azinphos Ethyl** | 16.7 | - | 41.2 | 0.03-3.4[o] | |
| | | **Chlorfenvinphos** | 16.7 | - | 19.7 | 1.4-3.9[o] | |
| | | **Diazinon** | 33.3 | 46.5 | 36 | 0.5-9.5[c], 2.3-132.3[i], 0.8-785[j], 8.4[l], 0.1-20.4[m], 17-34[n], 0.5-35.8[o], 0.06[q] | 53.6 (inf)[h] 281 (effl)[h] |
| | | **Malaoxon (MOX)** | 16.7 | - | 6.6 | 0.15-1.44[c] | |
| | **Organothiophosphates** | **Fenthion Oxon** | 16.7 | - | 12.2 | 2.64[m] | |
| | | **Fenthion Sulfoxide** | 16.7 | - | 10.8 | 2.64[m] | |
| | **Neo-nicotinoids** | **Imidacloprid** | 33.3 | 72.8 | 56.9 | 1.6-14.9[m], 0.31-66.5[o] | |

*For those pesticides detected only in one sample out of six (FD of 16.7%), maximum value and average value coincide. Standard deviation is included for those compounds with FD>50% (n≥3). Referenced values are given as average concentrations or average concentration ranges; a: Loos et al., 2010; b: Loos et al., 2009; c: Palma et al., 2014; d: Woudneh et al., 2007; e: Rodil et al., 2012; f: Kock et al., 2010; g: Rodríguez-Mozaz et al., 2015; h: Kock et al., 2013; i: Kock et al., 2012; j: Ricart et al., 2010; k: Rodriguez-Mozaz et al., 2004; l: Kampioti et al., 2005; m: Ccanccapa et al., 2016; n: Kock et al., 2011; o: Masià et al., 2013; p: Quintana et al., 2001; q: Palma et al., 2015; r: Proia et al. 2013

**Table 5.** Limits of detection, quantification and determination for those pesticides detected in the agricultural runoff.

| | PESTICIDES | LOD (ng L$^{-1}$) | LOQ (ng L$^{-1}$) | Limit of Determination (ng L$^{-1}$) |
|---|---|---|---|---|
| Acidic | 2,4- D | 1.4 | 4.5 | 7.5 |
| | MCPA | 2.4 | 8.1 | 8.1 |
| Chloroacetanilides | Alachlor | 1.3 | 4.4 | 4.4 |
| Phenyl Ureas | Diuron | 0.22 | 0.7 | 0.7 |
| | Linuron | 3.19 | 11 | 11 |
| Triazines | Cybutrine | 1.3 | 4.2 | 4.2 |
| | Desisopropyl atrazine | 19 | 63 | 63 |
| | Terbuthylazine | 0.2 | 0.8 | 1.1 |
| | Terbutryn | 0.1 | 0.3 | 0.3 |
| Organophosphates | Azinphos Ethyl | 6.7 | 22 | 22 |
| | Chlorfenvinphos | 0.1 | 0.5 | 0.5 |
| | Diazinon | 0.1 | 0.2 | 0.2 |
| | Malaoxon | 0.1 | 0.5 | 0.5 |
| Organ thiophosphates | Fenthion Oxon | 0.2 | 0.57 | 0.6 |
| | Fenthion Sulfoxide | 1 | 3.2 | 3.2 |
| Neonicotinoids | Imidacloprid | 1.8 | 6 | 6 |

The chloroacetanilide alachlor was detected at a concentration of 6 ng L$^{-1}$ in only one sample. Alachlor belongs to the list of priority substances subjected to regulation in surface waters, but the concentration found in our study is far below the maximum allowable concentration (MAC) of 700 ng L$^{-1}$ established for it in the Directive 2013/39/EC.

As for the phenylureas detected, diuron was present in all samples but at lower levels than the ubiquitous MCPA, ranging from 29-61 ng L$^{-1}$, whereas linuron was detected only in one sample (13 ng L$^{-1}$). For diuron, the concentrations found are in accordance with those previously reported in the surface water of the nearby Llobregat River basin (Köck-Schulmeyer et al., 2012; Proia et al., 2013; Ricart et al., 2010). A noticeable increase in the occurrence of diuron from 2000 to 2010 in this area has been reported, with maximum levels detected increasing from 9.5 ng L$^{-1}$ in 2003 (Kampioti et al., 2005) to 99 ng L$^{-1}$ in 2005-2006 (Ricart et al., 2010) and 818 ng L$^{-1}$ in 2010 (Köck et al., 2012). Yet, average and median results are similar to those presented in this study (42 ng $L^{-1}$ average). Diuron has also been frequently detected in urban wastewaters in the area of study, at average concentrations of 93 ng $L^{-1}$ (Köck-Schulmeyer et al., 2013b). Diuron is classified as a priority substance in the EU (Directive 2013/39/EU), with a maximum allowable concentration (MAC) in inland surface waters of 1.8 µg $L^{-1}$.

Regarding triazines, four out of the nine targeted compounds were present in the agricultural run-off, but only terbutryn was detected in all samples, at levels in the range 15-70 ng $L^{-1}$. Terbutryn has been previously detected in surface water of the Ebro River basin at similar concentrations, in the range of 0.9-31 ng $L^{-1}$, but with a lower frequency of detection of barely 25% (Ccanccapa et al., 2016a). Terbutryn is also included in the EU list of potential endocrine disruptors within category 1 (substances for which endocrine activity has been documented in one or more studies against a given living organism). These substances are given the highest priority for further research. It is also included in the EU priority list, and the MAC in inland surface waters was set to 0.34 µg $L^{-1}$. Three other triazines were also detected, but only in one sample (cybutrine, desisopropyl atrazine, and terbuthylazine), at concentrations of 14, 64 and 11 ng $L^{-1}$. Cybutrine is included in the list of priority substances with a MAC of 16 ng $L^{-1}$ in inland surface waters, a concentration only slightly higher than that measured in our agricultural runoff; on the other hand, terbuthylazine is included in the list of specific pollutants subject to regulation in Spain (RD 817/2015), with a MAC of 1000 ng $L^{-1}$ in inland surface waters, which is far higher than the 11 ng $L^{-1}$ found in our study.

Four organophosphates were detected (azinphos ethyl, chlorfenvinphos, diazinon, and malaoxon), but again only in one out of the six days of sampling, except for diazinon, which was detected two days (26 and 47 ng $L^{-1}$). Diazinon has been frequently detected in the same area (Köck-Schulmeyer et al., 2012; Proia et al., 2013; Ricart et al., 2010), and also in the Ebro River basin at maximum concentrations of 14 and 20 ng $L^{-1}$ in 2010 and 2011, respectively, but again with a low frequency of detection (25%) (Ccanccapa et al., 2016a). The sale of diazinon was forbidden in the EU in June 2008, following the Decision 2007/393/EC related to the Directive 91/414/EEC, which classified this pesticide as a non-authorized substance. This restriction explains its lower frequency of detection in more recent studies. Nevertheless, it was still detected in urban wastewaters in the range 479-607 ng $L^{-1}$ in 2009 (Rodriguez-Mozaz et al., 2015), and at similar levels in the following campaigns (Köck-Schulmeyer et al., 2013b).

The fenthion metabolites fenthion oxon and fenthion sulfoxide were detected in a single sample and at low concentrations (12 and 11 ng $L^{-1}$, respectively).

Finally, the neonicotinoid imidacloprid was detected two days of sampling, at concentrations of 41 and 73 ng $L^{-1}$. Imidacloprid is included in the EU Watch List (Commission Implementing Decision EC/2018/840), with a maximum acceptable method detection limit, established accordingly to the PNEC of neonicotinoids in water, of 9 ng $L^{-1}$; this means that the levels measured in the present study could represent a risk. Lower concentrations were detected in the nearby Ebro River basin in 2010-2011 (1.06 and 1.66 ng $L^{-1}$ average values, respectively) and with low detection frequencies (25%) (Ccanccapa et al., 2016). The highest individual contributions during the 6 days of sampling corresponded to MCPA, diuron, terbutryn and imidacloprid. Out of the four most abundant pesticides, terbutryn is the only one that is no longer authorized for use as a pesticide in the EU. However, it is yet widely applied as an algaecide/preservative in the coatings industry (Musgrave et al., 2011).

### 3.3. Removal of pesticides in the PBR

Ten out of the sixteen pesticides present in the agricultural run-off were fully eliminated (alachlor, linuron, cybutrine, deisopropyl atrazine, terbuthylazine, azynphos ethyl, chlorfenvinphos, malaoxon, fenthion oxon, and fenthion sulfoxide), but eight of them (all but cybutrine), as explained in the previous section, were present in the agricultural run-off only one day of sampling (Figure 5). The remaining 6 pesticides (2.4-D, MCPA, diuron, terbutryn, diazinon, and imidacloprid) were still detected in the mixed liquor of the PBR (effluent samples), meaning either incomplete removal or formation during the treatment process (higher concentration in the effluent than in the influent, i.e., negative removal) (see Table 6). Different removal mechanisms take place within microalgae-based systems, and their prevalence and relevance are directly linked to the physical-chemical properties of the studied contaminants. These properties determine their adsorption, bioaccumulation capacity, and biodegradation within the system. For instance, a higher water solubility implies a higher occurrence of the pesticide in the aqueous phase and higher mobility, hence a higher bioavailability for biodegradation, especially under long HRTs (days) (Blum et al., 2018). It can also imply a higher exposure to chemical or photochemical degradation. Pesticides are usually classified according to their log $K_{ow}$ as compounds having a low (0-1), moderate (1–3), or high (>3) potential to bioaccumulate. However, binding mechanisms involved in the sorption of these and other organic contaminants onto biomass (and other solid matrices), depend also on other factors apart from its $K_{ow}$ value. Different interactions between the functional groups of the contaminant and the biomass define the strength of the sorption, so parameters such as pH and ionic strength should not be neglected (Svahn and Björklund, 2015; Wu et al., 2015).

**Table 6.** Concentrations (ng $L^{-1}$) of the 16 different pesticides detected in the agricultural run-off (PBR influent) and the mixed liquor (PBR effluent) on every sampling day. AR: agricultural run-off. Values in italics indicate higher concentrations in the effluent than in the agricultural run-off.

| | | 12-7 | | 13-7 | | 14-7 | | 18-7 | | 19-7 | | 20-7 | |
|---|---|---|---|---|---|---|---|---|---|---|---|---|---|
| **CHEMICAL FAMILY** | **PESTICIDE** | **Inf** | **Eff** | **Inf** | **Eff** | **Inf** | **Eff** | **Inf** | **Eff** | **Inf** | **Eff** | **Inf** | **Eff** |
| **Acidic** | **2,4- D** | n.d. | n.d. | 23.2 | ***55.8*** | n.d. | n.d. | n.d. | n.d. | 9.7 | n.d. | <L.det | n.d. |
| | **MCPA** | 48.8 | n.d. | 45.1 | n.d. | 392.3 | n.d. | 108.7 | n.d. | 80.1 | 45.9 | n.d. | n.d. |
| **Chloroacetanilides** | **Alachlor** | 6 | n.d. | n.d. | n.d. | n.d. | n.d. | n.d. | n.d. | n.d. | n.d. | n.d. | n.d. |
| **Phenyl Ureas** | **Diuron** | 43.8 | ***52.3*** | 33.4 | ***57.2*** | 44.8 | ***61.1*** | 29.1 | ***40.9*** | 39.2 | ***45.3*** | 60.6 | ***71.1*** |
| | **Linuron** | n.d. | n.q. | n.d. | n.d. | n.d. | n.q. | 13.2 | n.d. | n.d. | ***71.7*** | n.d. | n.d. |
| **Triazines** | **Cybutrine** | 14.2 | n.d. | n.d. | n.d. | <L.det | n.q. | n.d. | n.d. | n.d. | n.d. | n.q. | n.d. |
| | **Desisopropyl atrazine** | n.d. | n.d. | n.d. | n.d. | 64.2 | n.d. | n.d. | n.d. | n.d. | n.d. | n.d. | n.d. |
| | **Terbuthylazine** | 11.3 | n.d. | n.d. | n.d. | n.d. | n.d. | n.d. | n.d. | n.d. | n.d. | n.d. | n.d. |
| | **Terbutryn** | 32.8 | ***61.4*** | 15.3 | ***68.9*** | 32.9 | ***72.5*** | 22.6 | ***88.5*** | 28.7 | ***94.1*** | 70 | ***99.5*** |
| **Organophosphates** | **Azinphos Ethyl** | 41.2 | n.d. | n.d. | n.d. | n.d. | n.d. | n.d. | n.d. | n.d. | n.d. | n.d. | n.d. |
| | **Chlorfenvinphos** | 19.7 | n.d. | n.d. | n.d. | n.d. | n.d. | n.d. | n.d. | n.d. | n.d. | n.d. | n.d. |
| | **Diazinon** | 25.5 | ***46.1*** | n.d. | ***35.2*** | n.d. | ***19.1*** | n.d. | ***14.8*** | n.d. | ***14.5*** | 46.5 | n.d. |
| | **Malaoxon** | 6.6 | n.d. | n.d. | n.d. | n.d. | n.d. | n.d. | n.d. | n.d. | n.d. | n.d. | n.d. |
| **Organ thiophosphates** | **Fenthion Oxon** | 12.2 | n.d. | n.d. | n.d. | n.d. | n.d. | n.d. | n.d. | n.d. | n.d. | n.d. | n.d. |
| | **Fenthion Sulfoxide** | 10.8 | n.d. | n.d. | n.d. | n.q. | n.d. | n.d. | n.d. | n.d. | n.d. | n.d. | n.d. |
| **Neonicotinoids** | **Imidacloprid** | n.d. | ***108.3*** | 72.8 | ***174.2*** | 41.1 | ***176.6*** | n.d. | ***67.9*** | n.d. | n.d. | n.d. | ***137.4*** |
| | **Cumulative Concentration** | **273** | **268.1** | **189.8** | **391.2** | **575.3** | **329.2** | **173.5** | **212.1** | **157.7** | **271.5** | **177.1** | **307.9** |

MCPA was efficiently removed (89% on average), showing complete elimination in 4 out of the 5 days it was found in the agricultural run-off. Its low $K_{ow}$ (-0.81) and high solubility in water (29.390 mg $L^{-1}$) (see Table 2) make both photodegradation and biodegradation the most feasible removal pathways within the PBR, discarding adsorption onto biomass. Indeed, MCPA has not been detected in sediments in any of the previous occurrence studies on this matrix (Köck-Schulmeyer et al., 2013a; Palma et al., 2015), except for the work of Ricart et al., (2010), in which MCPA was detected at trace concentrations (0.46-1.96 ng $g^{-1}$). It has not been detected either in aquatic organisms (shellfish, fish) (Barbieri et al., 2019; Köck et al., 2010). Its removal in WWTPs is contradictory, going from removals > 70% (Rodriguez-Mozaz et al., 2015) to the finding of higher concentrations in the secondary effluent than in the influent (Köck-Schulmeyer et al., 2013b). The other acidic herbicide detected, 2,4-D, with very similar physical-chemical properties as MCPA, showed also variable RE% of 100% and -140% in the two days it was detected in the PBR. According to the presented results, 2,4-D seems to be more resilient than MCPA. However, in WWTPs better removals have been reported for 2,4-D than for MCPA were reported by (Köck-Schulmeyer et al., 2013b).

The phenyl urea diuron was quantified in all influent and effluent samples at concentrations always higher in the mixed liquor than in the corresponding agricultural run-off samples. Its low water solubility (813 mg $L^{-1}$) and moderate $K_{ow}$ (2.87) make it prone to adsorb onto the biomass. Diuron is indeed one of the pesticides most commonly found in sediments (Köck-Schulmeyer et al., 2013a; Masiá et al., 2015, 2013; Palma et al., 2015; Ricart et al., 2010). Furthermore, it is very stable in soils and shows a low tendency to photodegrade. In WWTPs it has been frequently measured in the treated effluent whereas it had not been found in the corresponding inlet water(Köck-Schulmeyer et al., 2013b; Rodriguez-Mozaz et al., 2015). The concentrations of diuron registered in the effluent samples were always below its MAC according to Directive 2013/39/EU (1.8 µg $L^{-1}$).

The triazine terbutryn, present in all the investigated agricultural run-off samples, was also consistently found at higher concentrations in the mixed liquor than in the influent run-off throughout the experiment. This compound has a high $K_{ow}$ (3.66) and a low water solubility (25 mg $L^{-1}$), which indicates a high adsorption potential, put into evidence in previous works that reported its occurrence both in sediments and fish tissue (Barbieri et al., 2019; Barbieri et al., 2020; Ccanccapa et al., 2016; Masiá et al., 2015, 2013). During summer, higher photosynthesis activity leads to a higher pH in the mixed liquor, which could promote the desorption of terbutryn from the biomass ($Pk_a$ of 4.3 for terbutryn). Furthermore, the turbulent flow within the tubes could promote desorption from the biomass (Vassalle et al., n.d.). This desorption could be a feasible explanation for the higher concentrations measured in the mixed liquor of the PBR. Nevertheless, and similarly to diuron, concentrations registered in the effluent were never superior to the MAC established for this compound in the Directive 2013/39/EU (0.34 µg $L^{-1}$). Imidacloprid, alike diuron and terbutryn, was again always found at higher concentrations in the effluent (6 positive samples) than in the influent (2 positive samples), indicating that it is neither biodegraded/adsorbed by the microalgae/bacteria consortium in the PBR nor photodegraded. This pesticide shows a low $K_{ow}$ (0.57) and moderate water solubility (6.719 mg $L^{-1}$), so it tends to be present in the liquid phase. The LOQ estimated in the influent water is quite low (6 ng $L^{-1}$), so analytical issues (signal decrease due to matrix effects) to justify these results could be disregarded. Its presence as a conjugate in the agricultural run-off and its later de-conjugation during treatment could be a feasible explanation for the higher concentrations measured in the PBR effluent, but further research should be carried out for confirmation.

In a recent prioritization study conducted by von der Ohe et al., (2011) to point out the most problematic organic contaminants present in the European rivers, 73 contaminants, out of a total of 500 evaluated, were classified as potential priority pollutants in freshwater ecosystems, and among them were diazinon, diuron, terbutryn and MCPA. Overall, negative removals found in this system and other studies after CAS wastewater treatments could be partially explained by the transformation of pesticide metabolites present in the agricultural run-off or influent wastewaters (not monitored with this targeted analytical approach) into their original compounds. This back-transformation has been observed for different human metabolites of pharmaceuticals during wastewater treatment (García-Galán et al., 2012; Vieno et al., 2007). However, this hypothesis has been seldom investigated in the case of pesticides, and deserves further attention.

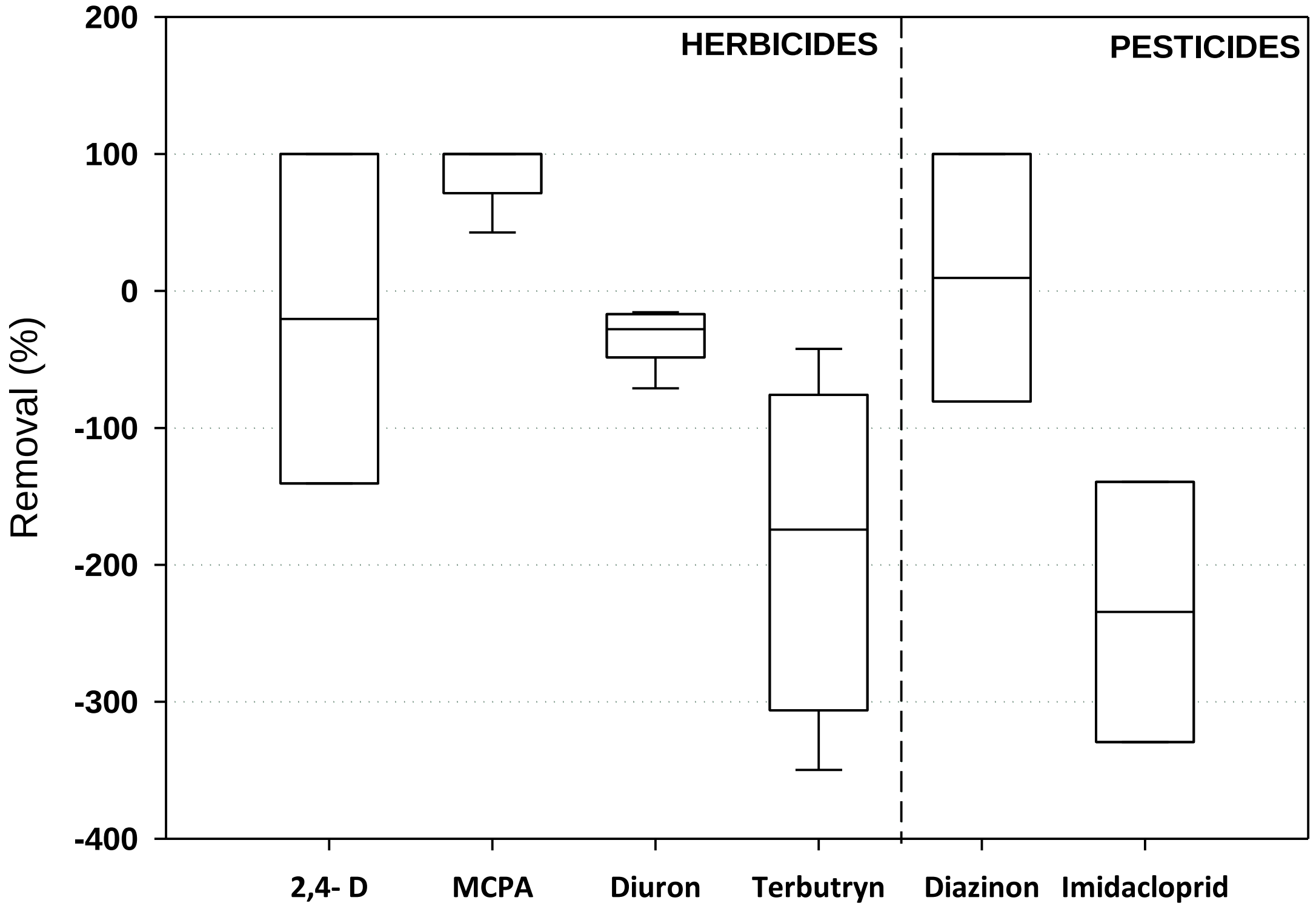

**Figure 5**. Removal rates (RE%) of the different pesticides studied. Only compounds with more than 2 RE% estimated have been included (alachlor, azinphos ethyl, chlorfenvinphos, desisopropil atrazine, fenthion oxon, fenthion sulfoxide, irgarol, linuron, malaoxon and therbuthylazine had RE%=100%, but n=1).

### 3.4. Risk assessment

HQs were calculated for those pesticides still present in the PBR effluent. The data used to calculate HQs, and the HQs obtained for the most conservative and medium scenarios (using the maximum and average measured concentrations) are summarized in Table 7. The PNEC values for freshwater coincide with European EQS or the homologs in the German water ecosystems protection policy (JD-UQN). In the case of imidacloprid, the PNEC corresponds to the maximum acceptable method detection limit set in the Watch List 2018/840.

According to the calculated HQ values, imidacloprid is the pesticide posing the highest risk to the aquatic ecosystem (HQ>10). This could be attributed to its very low PNEC value (8.3 ng $L^{-1}$). Furthermore, also driven by their low PNEC values (10-65 ng $L^{-1}$), a moderate risk could be expected from the presence of 2,4-D, terbutryn, and diazinon (1<HQ<10). MCPA would be the only pesticide for which no adverse effects could be expected in aquatic organisms (HQ<0.1), whereas for linuron, although the risk is low, potential adverse effects cannot be fully dismissed (0.1<HQ<1).

The overall hazard quotient of the PBR effluent samples ($HQ_{eff}$) was usually > 10 except for one day (Figure 6). This could be attributed to the relatively high concentrations of toxic pesticides detected in the PBR effluent samples, e.g., imidacloprid (68-177 ng $L^{-1}$), terbutryn (61-100 ng $L^{-1}$) and diuron (41-71ng $L^{-1}$). Further investigation should be carried out, as these pesticides were usually not present in the agricultural run-off or present at lower concentrations than in the effluent. The environmental risk posed by the discharged effluents is similar and constant throughout the treatment.

**Table 7.** Hazard quotients estimated for the pesticides detected in PBR effluent waters in different scenarios (maximum and average detected concentrations) and using PNEC value reported in freshwater ecosystems.

| **Pesticide** | **Maximum MEC (ng $L^{-1}$)** | **Average concentration (ng $L^{-1}$)** | **PNEC freshwater (µg $L^{-1}$)** | **$HQ_{max}$** | **$HQ_{average}$** |
|---|---|---|---|---|---|
| **2,4D** | 55.8 | 55.8 | 0.02* | **2.79** | 2.79 |
| **MCPA** | 45.9 | 45.9 | 0.5** | 0.09 | 0.09 |
| **Diuron** | 71 | 51.4 | 0.2** | **0.35** | 0.26 |
| **Linuron** | 71.7 | 71.7 | 0.1* | 0.72 | 0.72 |
| **Terbutryn** | 99.5 | 80.8 | 0.065 ** | **1.53** | 1.24 |
| **Diazinon** | 46.1 | 25.9 | 0.01* | 4.61 | 2.6 |
| **Imidacloprid** | 176.6 | 131.7 | 0.0083*** | **21.3** | 15.9 |

PNEC values obtained from the NORMAN Ecotoxicology Database (https://bit.ly/2Cm4zOE).
*:PNEC type: JDN-UQN;
**: PNEC-type: European EQS;
***: proposed EQS value, not in force.

It should be taken into account that the conventional risk assessment of chemical mixtures (especially for regulatory purposes) is usually based on the concentration addition for estimating the mixture toxicity (European Comission, 2009), ignoring synergistic toxicity, as well as additive effects or antagonistic effects (Baek et al., 2019). Different studies have demonstrated that synergistic interaction could be a relatively rare occurrence within realistic pesticide mixtures (low concentrations in mammals). However, indeed, studies devoted to chemical mixtures toxicities usually deal with less than 7 compounds (frequently binary mixtures), which is indeed unrealistic, and that synergistic toxicity may increase proportionally with the number of pollutants considered (Heys et al., 2016). Therefore, concentration addition may underestimate the overall risk posed by real environmental mixtures in complex matrices. This is out of the scope of the present study, but it is a hot topic within the scientific community, which is currently devoting a huge effort to discern and evaluate more realistic toxicity scenarios.

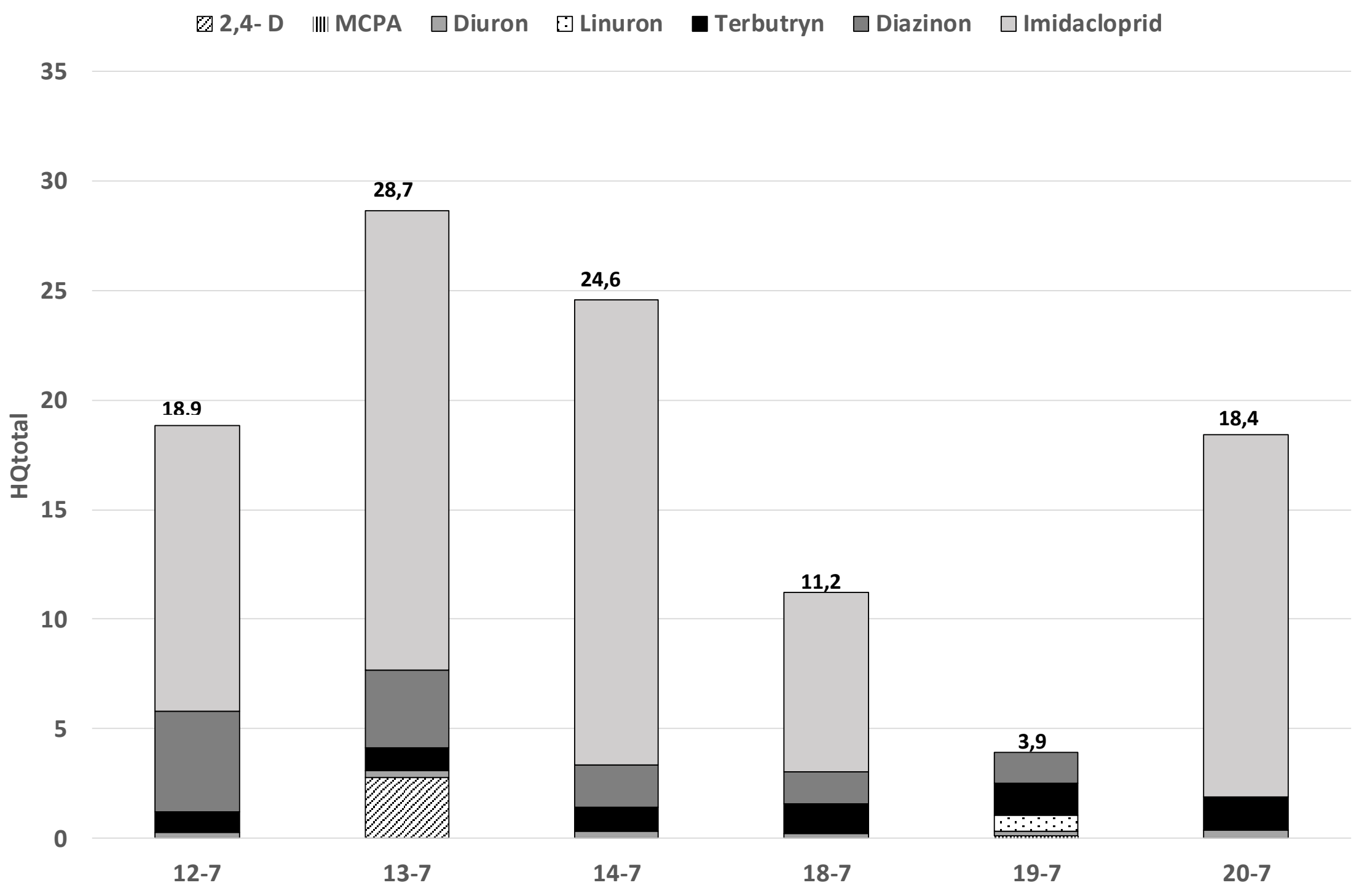

**Figure 6.** Cumulative HQ for each day of sampling and contribution of each pesticide to the theoretical total PBR effluent toxicity (values in bold are the daily sum of the individual HQs).

## 4. CONCLUSIONS

The performance of a semi-closed PBR to remove different types of pesticides has been evaluated. According to common physical-chemical parameters of the water, the PBR worked very efficiently to decontaminate agricultural run-off in terms of $NH_4^+$-N as well as most of the pesticides found in the original water. However, some important pesticides in terms of usage and toxicity, such as terbutryn, diuron, diazinon, and imidacloprid, were not efficiently removed or even appeared at higher concentrations in the effluent than in the influent of the PBR. The conversion of non-targeted pesticide transformation products or metabolites into the original compounds during PBR treatment could be a feasible explanation to these higher concentrations in the effluents, but it should be investigated in more detail. Similar results have frequently been observed in conventional WWTPs, which implies that the treatment capacity of WWTPs is not improved in this microalgae-based system. The satisfactory implementation of these systems is therefore reduced to areas where the most recalcitrant compounds are either little or not used. Terbutryn and diuron, listed as priority substances in the Water Framework Directive, should not be used as plant protection products in agriculture, and imidacloprid, included in the 2018 Watch List, may be subject to regulation shortly. Future research will aim to improve the overall efficiency of the PBR in terms of the elimination of pesticides, encompassing the testing and adjustment of relevant operational conditions, especially HRT and pH. The development of biofilm in the PBR system was unavoidable, despite the regular maintenance, and new strategies should be studied and set to tackle this drawback. All considered, nature-based systems can be determinant to achieve a more sustainable wastewater treatment scenario, due to the reduction of the energetic and economic operational costs. The quantification of these energy and economic savings could be specifically addressed in further research, as well as the environmental evaluation through life cycle assessment.

## 5. ACKNOWLEDGMENTS

This research was funded by European Union's Horizon 2020 research and innovation program within the framework of the INCOVER project (GA 689242) and WATERPROTECT project (GA-727450); by the Spanish Ministry of Science, Innovation, and Universities, the Research National Agency (AEI), and the European Regional Development Fund (FEDER) within the project AL4BIO (RTI2018-099495-B-C21), and by the Government of Catalonia (Consolidated Research Group 2017 SGR 01404-Water and Soil Quality Unit). M.J. García-Galán, E. Uggetti, and R. Díez-Montero would like to thank the Spanish Ministry of Economy and Competitiveness for their research grants (IJCI-2017-34601, RYC2018-025514-I, and FJCI-2016-30997, respectively).